\documentclass[aps,prl,twocolumn]{revtex4}
\usepackage{amssymb}
\usepackage{amsmath}
\usepackage{amsfonts}

\begin{document}

\title[Quasiclassical boundary resistance]{Revision of the 
quasiclassical boundary resistance of the metal/superconductor interface}
\author{N. Garc\'{\i}a,$^{1}$ L.R. Tagirov$^{1,2}$}
\affiliation{$^{1}$Laboratorio de F\'{\i}sica de Sistemas Peque\~{n}os 
y Nanotecnolog\'{\i}a, CSIC, 28006 Madrid, Spain \\
$^{2}$Kazan State University, 420008 Kazan, Russian Federation}


\begin{abstract}
Transparency of the metal/superconductor interface is discussed in a frame 
of the Kupryanov-Lukichev (K-L) quasiclassical boundary conditions. It is 
shown that the original K-L boundary transparency is expressed via the
single-channel Landauer conductance. It is a source of dramatic discrepancy
between the theory and experiments on the superconducting proximity effect.
The improved multi-channel derivation is proposed for the boundary
conditions, which rehabilitates consistency of the quasiclassical approach
to the boundary transparency problem, and eliminates at least an order of
magnitude of discrepancy between the theory and the experiment. The 
influence of the interface roughness and wave function symmetry mismatch 
at the boundary on the interface transparency is discussed. Warning is made 
to theories that use the quasiclassical boundary conditions with the 
interface transparencies close to unity.
\end{abstract}


\volumeyear{year}
\volumenumber{number}
\issuenumber{number}
\eid{identifier}
\date[A version of this paper is submitted to PRL 19 July 2005]{ }
\startpage{1}
\endpage{15}
\maketitle


Since works by Eilenberger \cite{Eilenberg}, Larkin and Ovchinnikov 
\cite{LO}, and Usadel \cite{Usadel} the quasiclassical theory is main 
calculation approach in physics of superconductivity. Most of calculations 
utilize the Usadel theory with the Kupriyanov-Lukichev (K-L) boundary 
conditions \cite{K-L} because the short mean-free path approximation 
matches most experiments with conventional superconductors 
(note that the short coherence-length, high-temperature 
superconductors refer mainly to a clean-limit case). As
long as the boundary resistance, which enters the K-L boundary conditions,
is considered as a phenomenological fitting parameter, the Usadel theory
gives adequate description of contact phenomena that has been demonstrated,
for example, for the Josephson effect by Golubov \textit{et al} 
\cite{Kupr} and for the proximity effect by Buzdin \cite{BuzdinRMP}. 
However, it was pointed out by the Salerno group 
\cite{Attanas1,Attanas2} that problems immediately arise 
when one tries to reproduce the best-fit
boundary resistance (or transparency of the interface), obtained from
proximity-effect experiments, making use of the physical parameters of 
materials in the contact. The authors employed the band offset (Fermi 
momenta mismatch) model for the quantum-mechanical transmission 
coefficient, 
\begin{equation}
D=\frac{4v_{1x}v_{2x}}{\left( v_{1x}+v_{2x}\right) ^{2}},  \label{eq1}
\end{equation}
which is plausible approximation as far as the free-electron model is used
as a background for the quasiclassical theory ($v_{ix}$ is a projection of 
the electron Fermi velocity of the $i$-th metal on the direction 
perpendicular to the interface). The quantum-mechanical transmission 
coefficient $D$ is related to the boundary resistance $R_{b}$ via parameter 
$\gamma _{b}$ as follows 
\begin{equation}
\gamma _{b}=\frac{\sigma _{2}AR_{b}}{\xi _{2}}=\frac{2l_{2}}{3\xi _{2}}
\left\langle \frac{x_{2}D}{1-D}\right\rangle ^{-1},  \label{eq2}
\end{equation}
where $A$ is an area of the contact, $\sigma _{2}=e^{2}p_{2}^{2}l_{2}/3\pi
^{2}\hbar ^{3}$ is the normal-phase conductivity, $\xi _{2}=\left( \hbar
D_{2}/2\pi T_{c}\right) ^{1/2}$ is the coherence length, $l_{2}$ and $p_{2}$
are the mean free path and the Fermi momentum of electrons, 
$D_{2}=v_{2}l_{2}/3$ is the conduction electron diffusion coefficient, 
$T_{c}$ is the superconducting transition temperature, and, finally, 
$x_{2}=\cos \theta _{2}$, where $\theta _{2}$ is the quasiparticle 
incidence angle. The subscript "2" refers to a metal \#2. 
The parameter $\gamma _{b}$ varies between zero
(virtual interface between identical metals) and physical infinity 
(an ultimate case of highly resistive insulating barrier at the interface). 
Following Aarts \textit{et al.} \cite{Aarts97} the Salerno group 
characterized the interface by the transparency parameter, 
\begin{equation}
\mathcal{T}=\frac{1}{1+\gamma _{b}},  \label{eq4}
\end{equation}
which varies between 0 and 1. From the fitting of their data on the 
proximity effect in Nb/Pd system \cite{Attanas1} the authors 
obtained $\mathcal{T}_{exp} $(Nb/Pd)$\simeq $0.46, 
whereas their estimation from Eqs. (\ref{eq1}) and 
(\ref{eq2}) gives $\mathcal{T}_{theor}$(Nb/Pd)$\simeq $0.98 using the Fermi
velocities $v_{Nb}$=2.73$\times $10$^{7}$ cm/s and $v_{Pd}$=2.00$\times 
$10$^{7}$ cm/s. For Nb/Cu and Nb/Ag couples the Salerno group obtained 
$\mathcal{T}_{exp}$(Nb/Cu)$\simeq $0.30 as compared with 
$\mathcal{T}_{theor}$(Nb/Cu)$\simeq $0.5 
($v_{Cu}$=1.57$\times $10$^{8}$ cm/s); 
and $\mathcal{T}_{exp}$(Nb/Ag)$\simeq $0.33 as compared with 
$\mathcal{T}_{theor}$(Nb/Ag)$\simeq $0.55 
($v_{Ag}$=1.39$\times $10$^{8}$ cm/s) \cite{Attanas2}. Similar 
low interface transparencies have been derived from proximity experiments on
Nb/Al couples: 
$\mathcal{T}_{exp}$(Nb/Al)$\simeq $0.2-0.25 
($\mathcal{T}_{theor}\simeq $0.8) \cite{Golubov95}, and Nb/CuNi couple: 
$\mathcal{T}_{exp} $(Nb/CuNi)$\simeq $0.2-0.35 
\cite{Fominov,Rusanov}. One may derive from the above data that there is 
\textit{systematic and drastic discrepancy} between 
the experimentally established transparency of the S/N
interface and the quasiclassical theory which is applied to reproduce the
transparency data. The discrepancy looks even more dramatic if we recognize
that one needs \textit{unrealistic} \textit{difference} in the free-electron
Fermi energies, one hundred times and more, to reproduce the boundary
resistance, say, Nb/Cu couple. The interface roughness or mismatch of 
the wave functions symmetry at the interface can not compensate so drastic 
misfit between the experiment and the theory. Then, one may come to a 
conclusion that there is global inconsistency in the quasiclassical approach 
when applied to description of the contact phenomena.

In this paper we show that source of the discrepancy is a very approximate
treatment of the boundary resistance, which has been made upon derivation of
the K-L boundary conditions. We develop further the K-L approach and propose
new derivation of the boundary conditions for the Usadel equations which
relaxes mainly the above mentioned discrepancy. We argue that approximation
of highly transparent interface (the quasiclassic superconducting pairing
function is continuous across the interface), which was oftenly used in
calculations of the contact phenomena, is hardly (if ever possible) to
realize in an experiment with conventional superconductors.

We start from the Zaitsev boundary conditions \cite{Zaitsev} for the
Eilenberger equations following Kupriyanov and Lukichev: 
\begin{equation}
g_{a1}=g_{a2}=g_{a},  \label{eq5}
\end{equation}
\begin{gather}
g_{a}\left\{R\left(1-g_{a}g_{a}\right) +\frac{D}{4}\left(
g_{s1}-g_{s2}\right) ^{2}\right\}  \notag \\
=\frac{D}{4}\left(g_{s1}-g_{s2}\right) \left(g_{s1}+g_{s2}\right) ,
\label{eq6}
\end{gather}
where $g_{s(a)i}$ are the symmetric (antisymmetric) Green functions, and 
$R=1-D$ is the reflection coefficient. The boundary condition (\ref{eq5})
ensures continuity of the charge current at the interface, while the second
one, Eq. (\ref{eq6}), relates the quasiparticles flux to a drop of their
density at the interface, and depends on transparency of the interface.
Using the explicit matrix structure of the Eilenberger functions \cite
{Zaitsev} close to the transition temperature $T_{c}$, we arrive at the
basic for our discussion, \textit{linearized} Zaitsev boundary condition
\begin{equation}
2Rf_{a}=D\left(f_{s2}-f_{s1}\right) ,  \label{eq8}
\end{equation}
where $f_{i}$ is the anomalous pairing Green function, which describes
superconductivity in the system.We have to notice here that equation 
(\ref{eq8}) has been derived \textit{without} any approximation 
concerning transparency of the interface. To proceed further we need also 
the linearized Eilenberger equations for the normal metal which read: 
\begin{equation}
l\cos \theta_2 \frac{df_{s2}}{dz}+f_{a2}=0,  \label{eq9}
\end{equation}
\begin{equation}
l\cos \theta_2 \frac{df_{a2}}{dz}+f_{s2}=\left\langle f_{s2}\right\rangle,
\label{eq10}
\end{equation}
where the angular brackets mean averaging over the solid angle. The original
logics of K-L derivation is as follows \cite{K-L} (see also review 
\cite{Lam-Rai}, Ch.~4): using a constraint, 
$\left\langle \cos\theta f_{a}\right\rangle =$const$(z)$, 
obtained after the solid-angle
averaging of equation (\ref{eq10}), and the antisymmetric function 
$f_{a2}\simeq -l_{2}\cos \theta_{2}(dF_{2}/dz)$, found from Eq. (\ref{eq9})
at $z\gg l_{2}$ in the lowest order on anisotropy, we obtain from Eq. 
(\ref{eq5}) the first linearized boundary condition at $z=0$,
\begin{equation}
l_{1}p_{1}^{2}\frac{dF_{1}}{dz}=l_{2}p_{2}^{2}\frac{dF_{2}}{dz},
\label{eq11}
\end{equation}
where $F_{i}=\left\langle f_{i}\right\rangle $ is the isotropic anomalous
Usadel function.

To derive the second boundary condition we divide at first both sides of 
equation (\ref{eq8}) by $2R$: 
\begin{equation}
f_{a}=\frac{D}{2R}\left( f_{s2}-f_{s1}\right) .  \label{eq12}
\end{equation}
Then, we substitute the approximate antisymmetric function, utilized above, 
for the left-hand side, replace the symmetric Eilenberger functions in the 
right-hand side by their Usadel counterparts, $f_{si}\longrightarrow F_{si}$, 
multiply (\ref{eq12}) by $\cos \theta _{2}$ and average over the angle theta. 
This procedure yields the second linearized boundary condition, 
\begin{equation}
\xi _{2}\gamma _{b}\frac{dF_{2}}{dz}=F_{1}-F_{2},  \label{eq13}
\end{equation}
with $\gamma _{b}$ given by Eq. (\ref{eq2}). Equations (\ref{eq11}) and 
(\ref{eq13}) are well known linearized Kupriyanov-Lukichev boundary
conditions \cite{K-L} for the Usadel functions. It has been shown
by Lambert \textit{et al} \cite{Lambert2} that they are valid only at
low transparencies of the S/N interface. Why does it so despite of 
the generic Zaitsev BC (\ref{eq8}) are valid for arbitrary transparency?

To answer the question let us have a look at the physical 
meaning of the linearized Zaitsev BC (\ref{eq12}) which 
is parental for the K-L one, Eq. (\ref{eq13}). It has a 
sense of the boundary condition for a \textit{particular trajectory} 
determined by the angles $\varphi $ and $\theta $. The language of 
trajectories is suitable when there is a continuous domain of quasiparticle 
incidence angles which accommodates macroscopic number of trajectories. 
In a case of a very few traversing trajectories it is more 
suitable to use a language of quantum-mechanical 
conductance channels. Then, the \textit{single-channel
Landauer conductance} \cite{Landauer} (in units of the conductance quantum 
$e^{2}/\pi h$) can be immediately recognized in the ratio $D/R$ which enters
the right-hand side of equation (\ref{eq12}). Subsequent derivation
procedure involves \textit{integration over the incidence angles} of the
single-channel Landauer conductance. Implication of the single-channel 
conductance to BC is crucial for understanding the source of 
insufficiency of the original K-L approximation reproduced above.

Indeed, the integrated single-channel Landauer conductance matches physical
expectation, that conductance of the interface is zero when the transparency
coefficient $D$ is zero (boundary resistivity is infinite). Alternatively,
the conductance is infinite (boundary resistivity is zero) when the 
contacting materials are identical, and the interface 
is virtual. On the other hand,
practically in all experiments with superconducting heterocontacts we have
many channels for conductance through the interface. From the physical
point of view one may expect that the \textit{multi-channel conductance} 
should enter the correct formulation of the boundary condition (\ref{eq13}). 
Then, we have to answer on a next question: is the angle-integrated, 
single-channel Landauer conductance represent correctly the multi-channel 
conductance? The answer \textquotedblleft no\textquotedblright\ is given 
at the end of the Section II of the paper by B\"{u}ttiker, Imry, Landauer 
and Pinhas \cite{BILP} (hereafter will be referred as BILP). 
In the Section IV of their paper the authors derive the multi-channel 
conductivity of the interface, valid at arbitrary transparency: 
\begin{equation}
G=\frac{2e^{2}}{\pi \hbar}\left( \frac{\sum\limits_{i=1}^{N^{\prime }}
D_{i}}{1+g_{l}^{-1}\sum\limits_{i=1}^{N}\dfrac{R_{i}}{v_{i}^{l}}
-g_{r}^{-1}\sum\limits_{i=1}^{N^{\prime }}\dfrac{D_{i}}{v_{i}^{r}}}
\right) ,  \label{eq14}
\end{equation}
where 
\begin{equation}
g_{l}=\sum\limits_{i=1}^{N}\left(v_{i}^{l}\right)^{-1}
,\,\,\,\,\,\,\,g_{r}=\sum\limits_{i=1}^{N^{\prime }}\left(v_{i}^{r}
\right) ^{-1} , \label{eq15}
\end{equation}
and $v_{i}$ are velocities of quasiparticles moving to the left (\textit{l})
and to the right (\textit{r}) over \textit{N} and \textit{N'} channels of
conductance, respectively. For a flat interface of macroscopic area the
summations over the conductance channels can be replaced by integration over
the incidence angles \cite{G-S}: $\sum
_{i=1}^{N_{\alpha }}\rightarrow \int\limits_{SS}{d}\varphi _{\alpha
}d\theta _{\alpha }\sin \theta _{\alpha }$, and the integration runs over
the semi-sphere (\textit{SS}) in the direction of the particles velocity.
Dependence of the multi-channel conductance (\ref{eq14}) on the
Fermi-energies of contacting metals and roughness of the interface between
them has been analyzed in detail by Garc\'{\i}a and Stoll \cite{G-S}.

To make evident relevance of a sheet conductance $\sigma _{\square }$
(conductance per unit square) of an interface for formulation of BC we
rewrite Eqs. (\ref{eq11}) and (\ref{eq13}) in terms of conserving fluxes: 
\begin{equation}
\sigma _{1}\frac{dF_{1}}{dz}=\sigma _{2}\frac{dF_{2}}{dz},
\label{eq17}
\end{equation}
\begin{equation}
\sigma _{2}\frac{dF_{2}}{dz}=\sigma _{\square }\left(F_{1}-F_{2}
\right) ,  \label{eq18}
\end{equation}
where 
\begin{equation}
\sigma _{\square }=\sigma _{\square }^{KL}=\left( \frac{e^{2}}{h}
\right) \left( \frac{p_{2}^{2}}{\pi ^{2}\hbar ^{2}}\right) \left(
2\pi \left\langle \frac{x_{2}D}{2(1-D)}\right\rangle
\right)   \label{eq19}
\end{equation}
is nothing else but the sheet conductance of the interface in the
Kupriyanov-Lukichev approximation. However, as we deduced from the above
analysis, the ultimate formulation of BC (\ref{eq18}) has to be expressed
via the multi-channel sheet conductance (at the current status, with the
last parentheses in Eq. (\ref{eq19}) replaced by the parentheses from Eq. 
(\ref{eq14})). Now, the main direction of logics is to find an analogue 
or a good approximation of the multichannel conductance remaining within 
the quasiclassical formalism.

As the first step let us look what will happen if we withdraw 
transformation of the original, particular trajectory Zaitsev 
BC (\ref{eq8}) into the single-channel BC form (\ref{eq12}). 
Starting from Eq. (\ref{eq8}) we replace: (1) the antisymmetric 
Eilenberger function $f_{a}$ in the left-hand side by the 
approximate expression used upon derivation of Eq. (\ref{eq11}); 
(2) the symmetric Eilenberger functions in the right-hand side 
by their Usadel counterparts. Then, we have in an output 
\begin{equation}
-2(1-D) l_{2}\cos \theta _{2}\frac{dF_{2}}{dz}=D\left(
F_{1}-F_{2}\right) .  \label{eq20}
\end{equation}
The procedure of the angular averaging described after Eq. (\ref{eq12}) 
yields BC (\ref{eq18}) with the sheet conductance 
\begin{equation}
\sigma _{\square}^{GT1}=\left(\frac{e^{2}}{h}\right) \left( \frac{
p_{2}^{2}}{\pi^{2}\hbar^{2}}\right) \left( 2\pi \left[\frac{
\left\langle x_{2}D\right\rangle}{2\left(1-3\left\langle x_{2}^{2}D
\right\rangle \right)}\right]\right) .  \label{eq22}
\end{equation}
The behavior of $\sigma _{\square}^{GT1}$ on a ratio 
of the Fermi energies $R_{F}=\varepsilon_{F2}/\varepsilon_{F1}\le 1$ 
is given by the dash line in Fig.~1. It is easy to see by comparison 
with Eq. (\ref{eq19}) that the main problem of K-L sheet conductance 
at high transparency of interface - the denominator, 
$1-D$, which is close to zero in the main domain of integration
angles - is relaxed already in Eq. (\ref{eq22}). However, the first 
step does not take into account  properly the influence of other 
conducting channels (or trajectories crossing the interface) on 
propagation of an electron along a particular trajectory. 

When there is a single conductance channel through the interface,
all other bulk conducting channels approaching the interface remain 
unperturbed. In our particular case of the S/N junction the 
superconducting order parameter parameter is flat near the 
impenetrable interface with the only an atomic-size hole which
accomodates one conductance channel. In the case of the transparent
interface, the other conductance channels induce mutual accomodation
of the pairing function wave function and the order parameter 
from the both sides  of the interface (see, for example, Fig.~3 
of Ref. \cite{BuzdinRMP}) creating gradients. Thus, the next 
step of approaching the true multichannel formulation could be
consideration of the Eilenberger function gradients. 

Careful look at the derivation procedure shows that upon 
replacement of the symmetric Eilenberger functions in the 
right-hand side of (\ref{eq8}) by their Usadel limits we have 
lost in Eq. (\ref{eq20}) gradient terms of the same order that we 
have in the left-hand side of this equation. 
In the dirty limit, the electron mean free path is the
shortest length compared with the superconducting or normal state 
coherence lengths. Then, in the spirit of Ref. \cite{V-T}, we 
expand the Eilenberger functions $f_{si}$ in powers of 
the mean free path along the particular trajectory, keeping 
the gradient terms, 
\begin{equation}
f_{si}\simeq \left\langle f_{si}\right\rangle +l_{zi}\frac{d\left\langle
f_{si}\right\rangle}{dz}=F_{i}+l_{zi}\frac{dF_{i}}{dz},
\label{eq23}
\end{equation}
and substitute them into the right-hand side of Eq. (\ref{eq8}). After the
angular averaging procedure we arrive at the second BC (\ref{eq18}) with 
the sheet conductance as follows: 
\begin{align}
\sigma _{\square }^{GT2}& =\left( \frac{e^{2}}{h}\right) \left(\frac{
p_{2}^{2}}{\pi ^{2}\hbar ^{2}}\right)   \notag \\
& \times \left( 2\pi \left[\frac{\left\langle x_{2}D\right\rangle}
{2\left( 1-\frac{3}{2}\left\langle x_{2}^{2}D\right\rangle -\frac{3}{
2}\left( \frac{p_{2}}{p_{1}}\right) ^{2}\left\langle x_{1}x_{2}D
\right\rangle \right) }\right] \right) .  \label{eq24}
\end{align}
Our sheet conductance (\ref{eq24}) is the next step to approximate the 
multichannel BILP sheet conductance (\ref{eq14}) in the gradient 
approximation. The two terms in the denominator of the big 
parentheses in (\ref{eq24}) are one-by-one 
counterparts of the similar terms, which we will have in 
(\ref{eq14}) after replacement of $R_{i}$ by $1-D_{i}$ in the denominator.
We believe (from considerations of energies involved in the transmission) 
that the result holds for the general, non-linear BC as well.
Dependence of the sheet conductance (\ref{eq24}) 
on $R_{F}$ is displayed by the thick solid line in Fig.~1. 
It can be seen that the GT2 curve goes much closer to the BILP sheet
conductance (dotted curve) than our first approximation given by 
Eq. (\ref{eq22}). The boundary conditions (\ref{eq17}) and (\ref{eq18}) 
can be expressed via the experimentally measured bulk resistivities of 
the contacting metals, $\rho _{1}=\sigma _{1}^{-1}$ and  
$\rho _{2}=\sigma_{2}^{-1}$, and the sheet resistance of the boundary, 
$AR_{b}=\sigma_{\square }^{-1}$.

We believe that our approach gives consistent quasiclassical approximation
to the multi-channel BILP sheet conductance of the interface. The 
improvement is drastic in the region of high transparencies: 
our sheet conductance $\sigma_{\square }^{GT2}$ is more than  
order of magnitude lower that the K-L one. Correspondingly, our 
sheet resistance of the interface, 
$AR_{b}^{GT}=\left(\sigma_{\square }^{GT2}\right)^{-1}$,
in the region of $R_{F}\lesssim 1$ is more than
order of magnitude higher than the K-L one. There are also other physical
reasons that may influence the boundary resistance. Garc\'{\i}a and Stoll 
have shown (Fig.~2 of Ref. \cite{G-S}) that the interface roughness
increases boundary resistance, i.e. $\gamma _{b}$. At the Fermi energy 
ratio $R_{F}=0.4$ this increase is about 20-60\% for different models 
of the interface roughness. The reduced overlapping and symmetry mismatch 
of the wave functions may also essentially suppress transparency of
the interface. Other physical aspect is that only the propagating
(dispersive) part $E_{prop}$ of the total electron energy, 
$E_{tot}=E_{prop}+E_{rot}$, computed in the free electron model, contributes
to the transport across the interface. So the narrow \textit{d}-electron
dispersion band may effectively produce reduced $R_{F}$ ratios as soon as
the localized rotational contribution $E_{prop}$ is subtracted from 
the total energy of the \textit{d}-state.

In a consequent way, one may think that light metals, for which the
approximation of free electrons is well justified, could be candidates for
the very transparent interface between them. If aluminum is chosen as a
superconductor ($\varepsilon _{F}($Al$)=11.7$ eV \cite{Ashk-Merm}) then the
series of noble metals as a counter-electrode (from $\varepsilon _{F}($Ag$
)=5.5$ eV to $\varepsilon _{F}($Cu$)=7$ eV) corresponds to $R_{F}=0.4-0.6$,
which implies essentially reduced interface transparency. It seems that
perfect interface transparency can not be realized in an experiment. 
In the view of the intrinsically reduced transparency of the interface
between any two metals we conclude that our quasiclassical boundary
resistance can be used in the full range of experimentally attainable
interface transparencies. On the other hand, all theories which 
essentially exploit perfect transparency of the interface have to 
be re-examined to check survival of predicted effects against the 
reduced transparency of the interface. 

The authors indebted to C.~Attanasio and S.~Prischepa for discussions
stimulating the study, and to M.~Kupriyanov, A.~Golubov, Yu.~Barash and
V.~Ryazanov for detailed discussion of the results. The support by the EC
grant NMP4-CT-2003-505282 is gratefully acknowledged, LRT acknowledge
partial support by the RFBR grant 03-02-17656.

\textbf{Figure Captions}

Fig. 1. Dependence of the normalized interface conductances on the Fermi
energy ratio $R_{F}$. The scale of the ordinate is logarithmic.

\end{document}